\documentclass[letter,apjl]{emulateapj-rtx4}

\usepackage{amsmath}
\usepackage{graphicx}
\usepackage{epsfig,psfrag,epic,eepic}
\usepackage{textcomp}
\usepackage{times}
\usepackage{enumerate}
\usepackage{amssymb, amsmath}
\usepackage{natbib}
\usepackage{color}
\usepackage{ulem}

\shorttitle{Inner Disk Component around HD~141569~A}
\shortauthors{Konishi et al.}

\newcommand{\osaka}{1}
\newcommand{\eureka}{2}
\newcommand{\goddard}{3}
\newcommand{\arizona}{4}
\newcommand{\washington}{5}
\newcommand{\charleston}{6}
\newcommand{\stisci}{7}
\newcommand{\mpia}{8}
\newcommand{\wyoming}{9}
\newcommand{\hst}{10}
\newcommand{\jpl}{11}
\newcommand{\tokyo}{12}
\newcommand{\oklahoma}{13}

\begin{document}

\title{Discovery of an Inner Disk Component around HD~141569~A\altaffilmark{$\star$}}

\author{Mihoko~Konishi\altaffilmark{\osaka, $\ast$}, 
	Carol~A.~Grady\altaffilmark{\eureka,\goddard}, 
	Glenn~Schneider\altaffilmark{\arizona}, 
	Hiroshi~Shibai\altaffilmark{\osaka}, 
    Michael~W.~McElwain\altaffilmark{\goddard},
	Erika~R.~Nesvold\altaffilmark{\washington}, 
	Marc~J.~Kuchner\altaffilmark{\goddard}, 
	Joseph~Carson\altaffilmark{\charleston}, 
	John.~H.~Debes\altaffilmark{\stisci}, 
	Andras~Gaspar\altaffilmark{\arizona}, 
	Thomas~K.~Henning\altaffilmark{\mpia}, 
	Dean~C.~Hines\altaffilmark{\stisci}, 
	Philip~M.~Hinz\altaffilmark{\arizona}, 
	Hannah~Jang-Condell\altaffilmark{\wyoming}, 
	Amaya~Moro-Mart\'in\altaffilmark{\stisci}, 
	Marshall~Perrin\altaffilmark{\stisci}, 
	Timothy~J.~Rodigas\altaffilmark{\washington,\hst}, 
	Eugene~Serabyn\altaffilmark{\jpl}, 
	Murray~D.~Silverstone\altaffilmark{\eureka}, 
	Christopher~C.~Stark\altaffilmark{\stisci}, 
	Motohide~Tamura\altaffilmark{\tokyo}, 
	Alycia~J.~Weinberger\altaffilmark{\washington}, 
	John.~P.~Wisniewski\altaffilmark{\oklahoma}
}

\altaffiltext{$\star$}{Based on data collected by the Hubble Space Telescope, operated by the Space Telescope Science Institute.}
\altaffiltext{\osaka}{Department of Earth and Space Science, Graduate School of Science, Osaka University, Osaka, Japan.}
\altaffiltext{$\ast$}{E-mail: konishi@iral.ess.sci.osaka-u.ac.jp.}
\altaffiltext{\eureka}{Eureka Scientific., Oakland, CA, USA}
\altaffiltext{\goddard}{Goddard Space Flight Center, Greenbelt, MD, USA.}
\altaffiltext{\arizona}{The University of Arizona, Tucson, AZ, USA.}
\altaffiltext{\washington}{Carnegie Institution of Washington, Washington, DC, USA.}
\altaffiltext{\charleston}{College of Charleston, Charleston, SC, USA.}
\altaffiltext{\stisci}{Space Telescope Science Institute, Baltimore, MD, USA.}
\altaffiltext{\mpia}{Max Planck Institute for Astronomy, Heidelberg, Germany.}
\altaffiltext{\wyoming}{University of Wyoming, Laramie, WY, USA.}
\altaffiltext{\hst}{Hubble Fellow}
\altaffiltext{\jpl}{Jet Propulsion Laboratory, California Institute of Technology, Pasadena CA, USA.}
\altaffiltext{\tokyo}{University of Tokyo, Tokyo, Japan.}
\altaffiltext{\oklahoma}{University of Oklahoma, Norman, OK, USA.}

\begin{abstract}\noindent
We report the discovery of a scattering component around the HD~141569~A circumstellar debris system, interior to the previously known inner ring. The discovered inner disk component, obtained in broadband optical light with HST/STIS coronagraphy, was imaged with an inner working angle of 0\farcs25, and can be traced from 0\farcs4 ($\sim$46~AU) to 1\farcs0 ($\sim$116~AU) after deprojection using $i$=55\arcdeg. The inner disk component is seen to forward scatter in a manner similar to the previously known rings, has a pericenter offset of $\sim$6~AU, and break points where the slope of the surface brightness changes. It also has a spiral arm trailing in the same sense as other spiral arms and arcs seen at larger stellocentric distances. The inner disk spatially overlaps with the previously reported warm gas disk seen in thermal emission. 
We detect no point sources within 2\arcsec ($\sim$232~AU), in particular in the gap between the inner disk component and the inner ring. Our upper limit of 9$\pm$3~M$_J$ is augmented by a new dynamical limit on single planetary mass bodies in the gap between the inner disk component and the inner ring of 1~M$_J$, which is broadly consistent with previous estimates.
\end{abstract}

\keywords{circumstellar matter --- stars: imaging --- stars: individual (HD~141569~A)}

\section{Introduction}
\label{intro}
There are a number of circumstellar disks associated with young stars which can be characterized as having small dust masses, but still retaining gas. Some show localized gas emission (e.g., $\beta$~Pic; \citealp{Dent+Wyatt+Roberge+etal_2014}), but others have centrally concentrated gas and mm-sized debris (e.g., HD~21997; \citealp{Moor+Juhasz+Kospal+etal_2013, Kospal+Moor+Juhasz+etal_2013}). There are the disks in which gaps or cavities have formed, possibly in tandem with giant planet formation. In particular, these systems challenge the practice of inferring the physical properties of the debris disks by modeling the infrared spectral energy distribution (SED).

HD~141569~A is a B9.5 star \citep{Merin+Montesinos+Eiroa+etal_2004} located at 116$\pm$8~pc \citep{vanLeeuwen_2007} from the Sun, and its age is reported to be 5$\pm$3~Myr \citep{Weinberger+Rivh+Becklin+etal_2000}. A mid-infrared excess detected by the \textit{Infrared Astronomical Satellite} (IRAS; \citealp{Walker+Wolstencroft_1988, Sylvester+Skinner+Barlow+etal_1996}) suggested that dust exists around HD~141569~A. Disk structures around HD~141569~A were first imaged by \citet{Augereau+Lagrange+Mouillet+etal_1999}, and \citet{Weinbergaer+Becklin+Schneider+etal_1999} discovered two rings using the Near Infrared Camera and Multi-Object Spectrometer (NICMOS) aboard \textit{Hubble Space Telescope} (HST). The system was subsequently imaged in the optical \citep{Mouillet+Lagrange+Augereaus+etal_2001,Clampin+Krist+Ardila+etal_2003} using the HST Space Telescope Imaging Spectrograph (STIS) and Advanced Camera for Surveys (ACS), respectively. The disk has also been imaged using ground-based adaptive optics systems \citep{Boccaletti+Augereau+Marchis+etal_2003,Janson+Brandt+Moro-Martin+etal_2013,Wahl+Metchev+Patel+etal_2014,Biller+Liu+Rice+etal_2015,Mazoyer+Boccaletti+Choquet_2016}. While complex structures were detected, the coronagraphic studies did not find dust interior to $\sim$175~AU due to analysis technique issues. However, mid-infrared observations detected thermal emission from small particles within 1\arcsec \ of the star \citep{Fisher+Telesco+Pina+etal_2000,Marsh+Silverstone+Beckelin+etal_2002}. These observations used filters containing strong polycyclic aromatic hydrocarbon (PAH) transitions, so that the nature of the material they detected remained unclear. Gas in the system has been studied in the millimeter \citep{Pericaud+DiFolco+Dutrey+etal_2014}, the mid- and far-infrared \citep{Thi+Pinte+Pantin+etal_2015}, with an inner disk component detected from $\sim$13~AU to $\sim$59~AU \citep{Goto+Usuda+Dullemond+etal_2006} using its distance of 116~pc. HD~141569~A has two M-type companions ($\sim$8\arcsec, \citealp{Weinberger+Rivh+Becklin+etal_2000}) which may produce the outer spirals \citep{Clampin+Krist+Ardila+etal_2003}.

Herein we present new imagery of the HD~141569~A circumstellar debris system. The HST/STIS imagery has an inner working angle (IWA) of 0\farcs25. Our observations confirm the previously reported outer disk structures and discover the presence of an inner dust disk component.

\section{Observations and Data Reduction}
\subsection{Observations}
We imaged HD~141569~A and a PSF (Point Spread Function) reference star (HD~135298, A0V, B-V=0.08)\footnote[14]{taken from SIMBAD, http://simbad.u-strasbg.fr/simbad/} using HST/STIS on 2015 June 12 and on 2015 August 18, as part of the general observer program (ID:13786, PI: G.~Schneider).  We observed using two coronagraphic image plane masks, WEDGEA1.0 and BAR5 (see \S 12.11 of the STIS Instrument Handbook; \citealp{STIS_inst_handbook}), and observed the PSF reference star interleaved with the HD 141569 observations (Table~\ref{table-log}). The WEDGEA1.0 imagery is used, following \citet{Schneider+Grady+Hines+etal_2014}, to provide large area deep imaging, at the cost of saturation near the coronagraphic wedge. The remaining data are shallower, were obtained using BAR5 (``the bent finger''), fill in the region saturated in the WEDGEA1.0 imagery and provide an IWA as small as 0\farcs25\footnote[15]{See ``Hubble Space Telescope STIS Coronagraphic BARs'',  http://www.stsci.edu/hst/stis/strategies/pushing/coronagraphic\_bars}. For the BAR5 observations, we employed three-point dithering (center and $\pm$0.25 pixel perpendicular to the long axis of the bar) to mitigate the effects of target mis-centering and to achieve the smallest possible IWA. Three sets of BAR5 data and one set of WEDGEA1.0 data were obtained per visit. The observations were divided into several sub-exposures per set to facilitate cosmic ray removal. The total integration time of HD~141569~A is 9.67$\times$10$^3$ (WEDGEA1.0) and 9.07$\times$10$^2$~seconds (BAR5), and that of PSF reference star is 3.22$\times$10$^3$ (WEDGEA1.0) and 1.09$\times$10$^2$~seconds (BAR5). We used sub-array readouts to reduce overhead times for our many exposures. The field of view (FoV) is 1024~pix$\times$427~pix (52\arcsec$\times$22\arcsec) and 1024~pix$\times$100~pix (52\arcsec$\times$5\farcs1) in WEDGEA1.0 and BAR5, respectively. The data were taken at 6 spacecraft roll angles (See Table~\ref{table-log} $ORIENTAT$). As demonstrated in \citet{Schneider+Grady+Hines+etal_2014}, combining such data reduces the area obscured by the STIS coronagraphic occulting structure and decorrelates image artifacts that are stable in the instrument. 
\subsection{Data reduction}
\label{reduc}
For basic instrumental calibration at the exposure level, we made use of the STIS calibration pipeline $calstis$ software \citep{STIS_data_handbook} and calibration reference files provided by the Space Telescope Science Institute (STScI). The $calstis$ pipeline software performs bias and dark current subtraction, as well as flat-field correction producing ``FTL'' files in instrumental count units. We median combined $NUMEXP$ exposures in each set (4 exposures in WEDGEA1.0 and 8 exposures in BAR5), in order to remove pixels affected by cosmic rays, and then converted the count unit to a count~s$^{-1}$ unit by dividing each exposure time ($EXPTIME$ in Table~\ref{table-log}). 

The stellar PSF was subtracted using the method described in \citet{Schneider+Grady+Hines+etal_2014}, using the Interactive Data Language (IDL)-based IDP3 (Image Display Paradigm \#3; \citealp{Stobie+Ferro_2006}) software\footnote[16]{https://archive.stsci.edu/prepds/laplace/idp3.html}. This step subtracts a scaled and registered version of the reference star HD~135298 from each HD~141569~A target image. The brightness scaling factor was calculated as 1.067 from the difference of $B$-magnitudes between the HD~141569~A (7.20) and HD~135298 (7.27)\footnotemark[14]. We used the $B$-magnitude in this calculation because the effective wavelength of this bandpass is most representative of a B9.5 star as seen by HST/STIS ($\lambda_{eff}\sim$4400\AA). 
The star's position was defined as the crossing point of the diffraction spikes, since the star was occulted by the WEDGEA1.0/BAR5 and was not measured directly. In the BAR5 observation, the PSF subtraction was done in all target-to-PSF template dither combinations.  We selected the best subtraction in each target image (top 2$\times$3 panels in Figure~\ref{fig-reduc}). For example, we made 9 subtracted images per visit (3 dithering target images $\times$ 3 dithering reference images), and then selected one per dithering (3 images in this case) which achieved the smallest IWA.  Within each visit, after target-minus-PSF template subtraction of the images in the instrument frame, we rotated the difference images about the occulted star to a common celestial frame. Then, suitable masks were made with IDP3 for each rotated image to occult non-physical features (diffraction spikes, coronagraphic mask structures, saturated pixels) and masking of the binary components HD~141569~BC. All images were registered to a common center and then median combined (bottom 3 panels in Figure~\ref{fig-reduc}). This combination of data from 6 different spacecraft roll angles removes all hot and dead pixels. In WEDGEA1.0 data, we also used the $cosmicrays$ task in Image Reduction and Analysis Facility (IRAF) before PSF subtraction to remove the hot and dead pixels, since there were not enough frames at each roll angle to adequately filter them. Finally, the BAR5 image was inserted in the void area of the WEDGEA1.0 image.

We did not use processing techniques for angular differential imaging (e.g., \citealp{Marois+Lafreniere+Doyon+etal_2006, Lafreniere+Marois+Doyon+etal_2007, Soummer+Pueyo+Larkin_2012}) because they introduce structure distortions \citep{Milli+Mouillet+Lagrange+etal_2012} and the BAR5 imagery has a limited amount of data.

\section{Results}

\subsection{Discovery of an inner disk component}
\begin{figure}
 \centering
 \includegraphics[width=\linewidth]{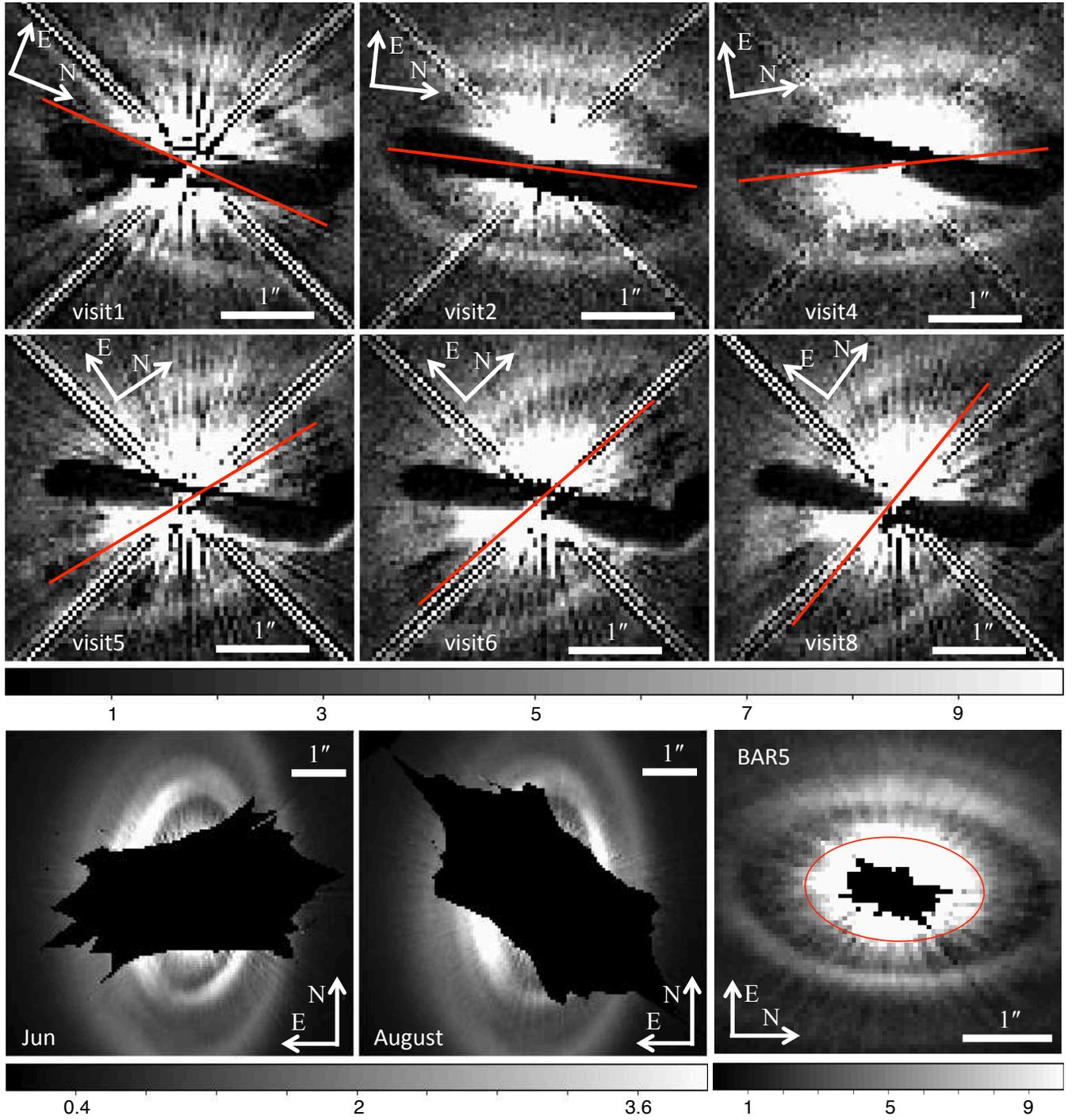}
 \caption{Top 2$\times$3 panels: PSF-template subtracted images of HD~141569 curcumstellar disk at six separate field orientations using the BAR5 mask revealing the inner dusk component (Component~C), and inner ring component (Component~B). Red lines shows major axis of Component~B. See Table~\ref{table-log} for detail. Bottom-left 2 panels: Deep imaging with the WEDGEA1.0 mask obscures most of Component~C imaged with BAR5, but cleanly images the rings (Component~A and B) beyond, as shown separately at the two observational epochs designed to provide maximum visibility around the star. Bottom-right panel: The merged BAR5 image with a red line represents a disk at $i$=55$\arcdeg$ and $PA$=357$\arcdeg$. The scale unit is count~s$^{-1}$.}
\label{fig-reduc}
\end{figure}

We detect an inner disk component interior to the previously reported rings in all 6 BAR5 observations and the outer extent in WEDGEA1.0 data (Figure~\ref{fig-reduc}). In BAR5 data (top panels of Figure~\ref{fig-reduc}), the inner signal is elliptical in shape, aligned with the disk major axis ($PA$=357\arcdeg \ from the north), and moves with the sky. This is not the behavior expected from PSF subtraction residuals which are very close to circularly symmetric at low spatial frequencies \citep{Schneider+Grady+Hines+etal_2014} and intrinsic to the disk due to the system inclination to the line of sight. We therefore conclude that we have discovered an inner component to HD~141569~A's disk. In the bottom-right panel of Figure~\ref{fig-reduc}, we display the merged BAR5 image with a elliptical line of $i$=55\arcdeg \citep{Mouillet+Lagrange+Augereaus+etal_2001}. Figure~\ref{fig-image} shows the merged WEDGEA1.0 and BAR5 imagery displayed with a stretch optimized for signal within 4$\arcsec$ of the star, as well as a cartoon showing our adopted nomenclature for the disk structures. We refer to the outer ring as Component~A, the inner ring as Component~B, and the newly imaged inner disk component as Component~C. We also refer to the gap between Component~A and B as Gap~AB, and the gap between Component~B and C as Gap~BC.

Component~C has no flux drop down to 0\farcs25, and an outer radial extent of 1\farcs0 defined from breaks seen in the radial surface brightness (SB) profile (see Section~\ref{sec-sbsd} in detail). The fit of the bottom-right panel of Figure~\ref{fig-reduc} follows Component~C and suggests it is also at $i$=55\arcdeg. We therefore use this specific angle to deproject the disk. In the surface density (SD) view, Component~C has a spiral arm at 1\farcs1 (see Section~\ref{sec-sbsd}). 

Figure~\ref{fig-image}~(a) shows a final composite image, with both the data void and M-type companions masked as black. Figure~\ref{fig-image}~(a) reproduces the previously reported azimuthal brightness asymmetries in Component~A and B \citep{Weinbergaer+Becklin+Schneider+etal_1999, Mouillet+Lagrange+Augereaus+etal_2001, Clampin+Krist+Ardila+etal_2003}, and suggests that these extend into Component~C. For further measurement and analysis, we deprojected the system using inclination $i$=55$\arcdeg$ to make Figure~\ref{fig-image}~(b), and then follow the \citet{Clampin+Krist+Ardila+etal_2003} methodology to correct the phase function and illumination effects. After deprojection 
(Figure~\ref{fig-image}~(b)), the stellocentric mask-limited IWA of our data is 0\farcs4 ($\sim$46~AU).

\subsection{Grain forward scattering}
We found a similar brightness asymmetry between the east and west for Component~C along the disk minor axis (see Figure~\ref{fig-image}~(a)) as previously reported for Component~A and B in the visible and near-infrared \citep{Weinbergaer+Becklin+Schneider+etal_1999,Mouillet+Lagrange+Augereaus+etal_2001,Clampin+Krist+Ardila+etal_2003}. The radial SB of the east side between 0\farcs7--1\farcs0 is 2.0$\pm$0.1$\times$ brighter than that of the west side (SB calculation is described in Section~\ref{sec-sbsd}). This is similar to the measured asymmetry in \citet{Mouillet+Lagrange+Augereaus+etal_2001} and slightly larger than \citet{Weinbergaer+Becklin+Schneider+etal_1999} at 1.1~$\mu$m. This asymmetric ratio corresponds to $g$=0.14--0.15 in the Henyey-Greenstein phase function \citep{Henyey+Greenstein_1941} assuming a geometrically thin disk and $i$=55\arcdeg. \citet{Clampin+Krist+Ardila+etal_2003} reported that Component~A and B are not azimuthally uniform.  We adjusted $g$ values assuming that this non-uniform SB feature is continued in Component~C and obtained a best fit value with $g$=0.1. This is consistent with the result of \citet{Weinbergaer+Becklin+Schneider+etal_1999} and \citet{Mouillet+Lagrange+Augereaus+etal_2001}. We note that our $g$ value is a lower limit because the degree of forward scattering for debris disks is underestimated \citep{Hedman+Stark_2015}. This problem was addressed in \citet{Stark+Schneider+Weinberger+etal_2014}. Figure~\ref{fig-image}~(c) shows the image after correcting for the scattering phase function asymmetry. 

\subsection{Radial surface density and brightness profile}
\label{sec-sbsd}
\begin{figure}
 \centering
  \includegraphics[width=\linewidth]{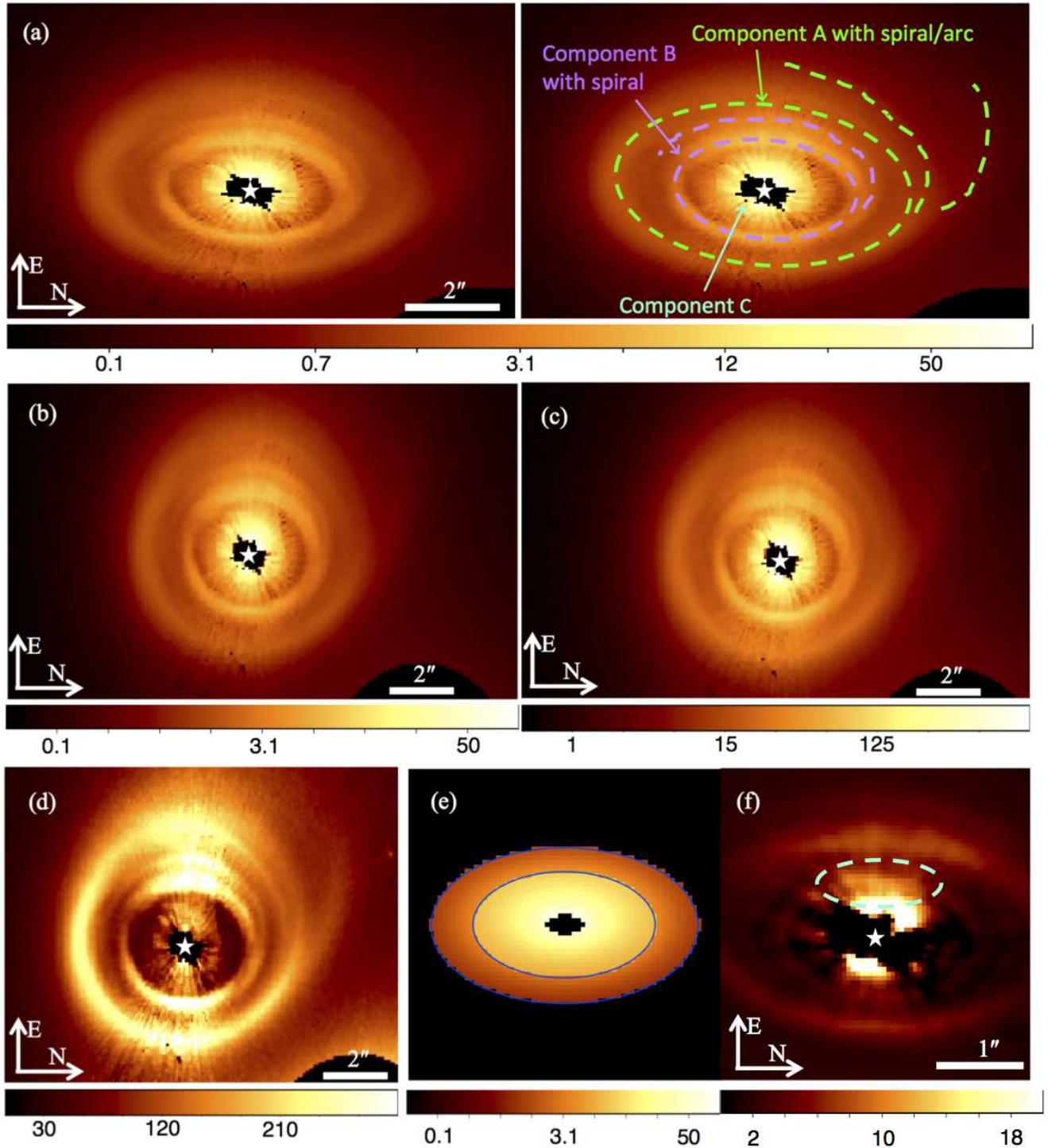}
 \caption{(a) Final image. In the right image, our nomenclature is annotated on the same image as presented at left. (b) Deprojected image assuming $i$=55\arcdeg. (c) Deprojected and phase corrected image using $g$=0.1. (d) Deprojected and phase corrected image multiplied by $r^2$, which is a proxy to a surface density map. Star marks are the primary star's position. The central obscuration and binary stars at bottom right are masked as black. (e) Toy disk model made in Section~\ref{sec-sbsd}. Blue ellipses are boundary of model disks. (f) Composite BAR5 image subtracted a toy disk. The color scale is lognormal in (a), (b), (c), and (e), and linear in (d) and (f). The scale unit is count~s$^{-1}$.} 
 \label{fig-image}
\end{figure}

\begin{figure}
 \centering
 \includegraphics[width=\linewidth]{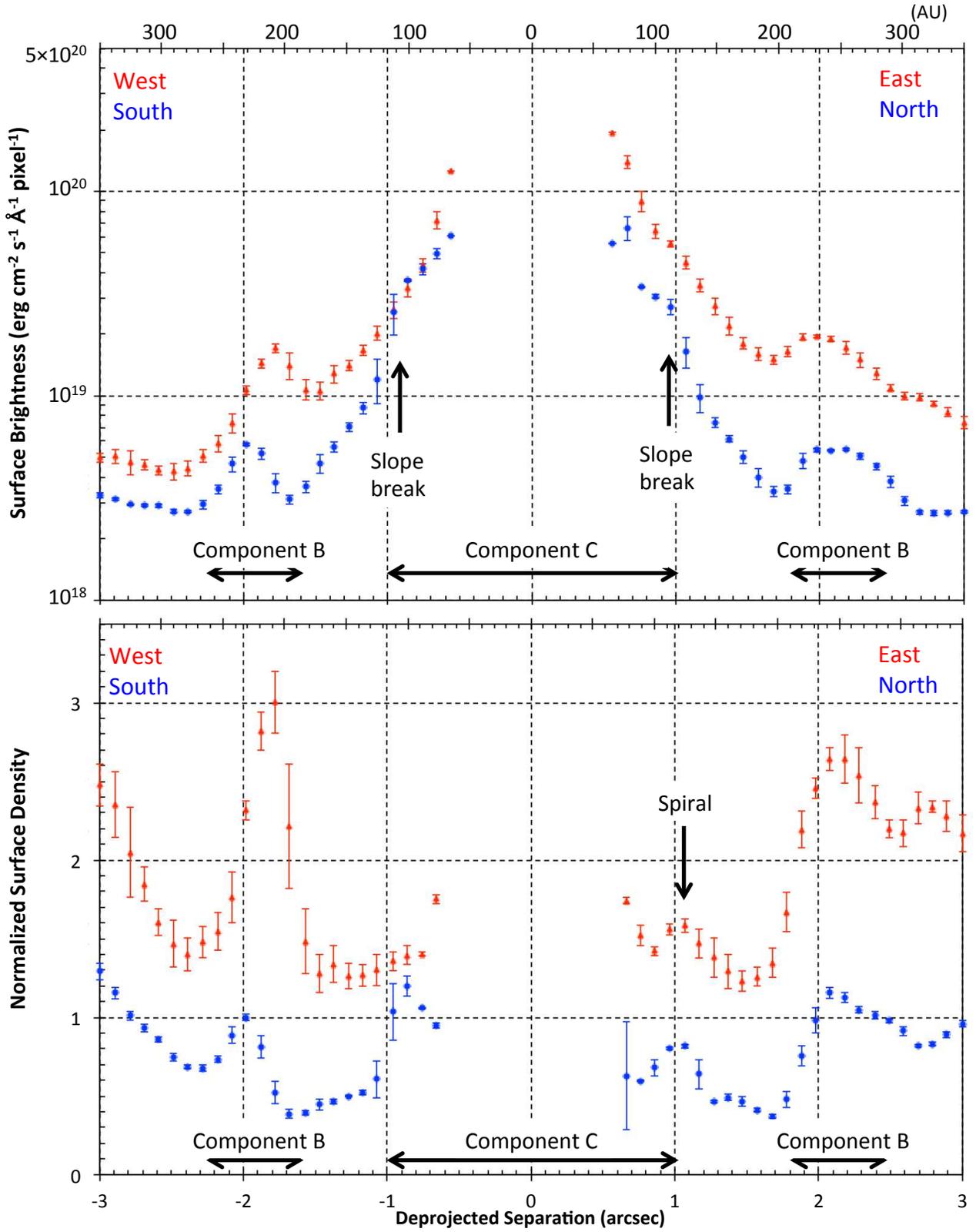}
 \caption{Top panel: Surface brightness (SB) profile along the major (N/S) and minor (E/W) axis of Figure~\ref{fig-image}~(b). Bottom panel: normalized surface density (SD) profile along the major and minor axis of Figure~\ref{fig-image}~(d). Plotted here is the normalized SD profile after correcting for a phase function assuming $g$=0.1 asymmetry parameter with $i=$55\arcdeg. 
 For normalization, the peak value of south of Component~B was used.}
\label{fig-sb}
\end{figure}

We multiplied Figure~\ref{fig-image}~(c) by $r^2$ (arcsec$^2$) to correct for the radial illumination, which provides a proxy SD map (see Figure~\ref{fig-image}~(d)). In order to investigate detailed structures, the SB and SD profiles were made (Figure~\ref{fig-sb}) using the deprojected image and the SD image (Figure~\ref{fig-image}~(b) and (d)). We calculated the mean within each small section (radial: 2~pixels, azimuth: 5\arcdeg), and converted counts~s$^{-1}$~pixel$^{-1}$ to flux density $F_{\lambda}$ (erg~cm$^{-2}$~s$^{-1}$~\AA$^{-1}$~pixel$^{-1}$) according to the equation described in \S~5.3 of \citet{STIS_data_handbook}, and errors on these measurements were calculated as the 1$\sigma$ scatter within a section. The SD profile in Figure~\ref{fig-sb} was normalized by the south peak of Component~B (2.9$\times$10$^{20}$~erg~arcsec$^2$~cm$^{-2}$~s$^{-1}$~\AA$^{-1}$). We note again that STIS has a broad bandpass and cannot specify a single wavelength. For STIS, 1~count~s$^{-1}$ is 4.55$\times$10$^{-7}$~Jy according to \citet{Schneider+Grady+Hines+etal_2014} and one square arcsec is approximately 388~pixels.

The SB of Component~C has slope breaks, which are points where the SB drops, along the major axis at $\sim$1\arcsec. The location of the breaks is 1\farcs0 ($\sim$116~AU) in the north and 0\farcs9 ($\sim$104~AU) in the south. The difference between the north and south breaks indicate that the Component~C center is offset of 0\farcs05 ($\sim$6~AU) toward the north. There is no clear break in the east and west side probably due to an elongation effect introduced by the deprojection.  We place no tight constraint on the Component~C center along the minor axis, but note that Gap~BC is slightly shifted to east according to SB in Figure~\ref{fig-sb}. It might be caused by a pericenter offset of Component~C. A spiral arm could be detected at 1\farcs1 ($\sim$128~AU) in the east side seen in Figure~\ref{fig-image}~(d) and SD profile in Figure~\ref{fig-sb}, with same sense as the outer spirals. 

We made a toy disk model (Figure~\ref{fig-image}~(e)) using SB indices, and subtracted it from the composite BAR5 image. The SB has two components inside Component~B, and thus the toy disk has a $-1.4$ index from 0\farcs25 to 1\farcs0 (Component~C) and a $-3.0$ index from 1\farcs0 to 1\farcs4 (halo between Component~B and C). We also assumed $i$=55\arcdeg, $g$=0.1, and $PA$=357\arcdeg, and pericenter offset of 1~pixel ($\sim$6~AU) toward north when made and subtracted it. Figure~\ref{fig-image}~(f) shows the 3$\times$3-box smoothed image. It has boundaries due to edges of the toy disk, but we detect brighter residuals in the east side extended at Component~C (enclosed area by a cyan line). These residuals may have a spiral arc-like geometry similar to other arcs in this system.

\subsection{Sensitivity to exoplanets}
\label{sec-dl}
\begin{figure}
 \centering
 \includegraphics[width=\linewidth]{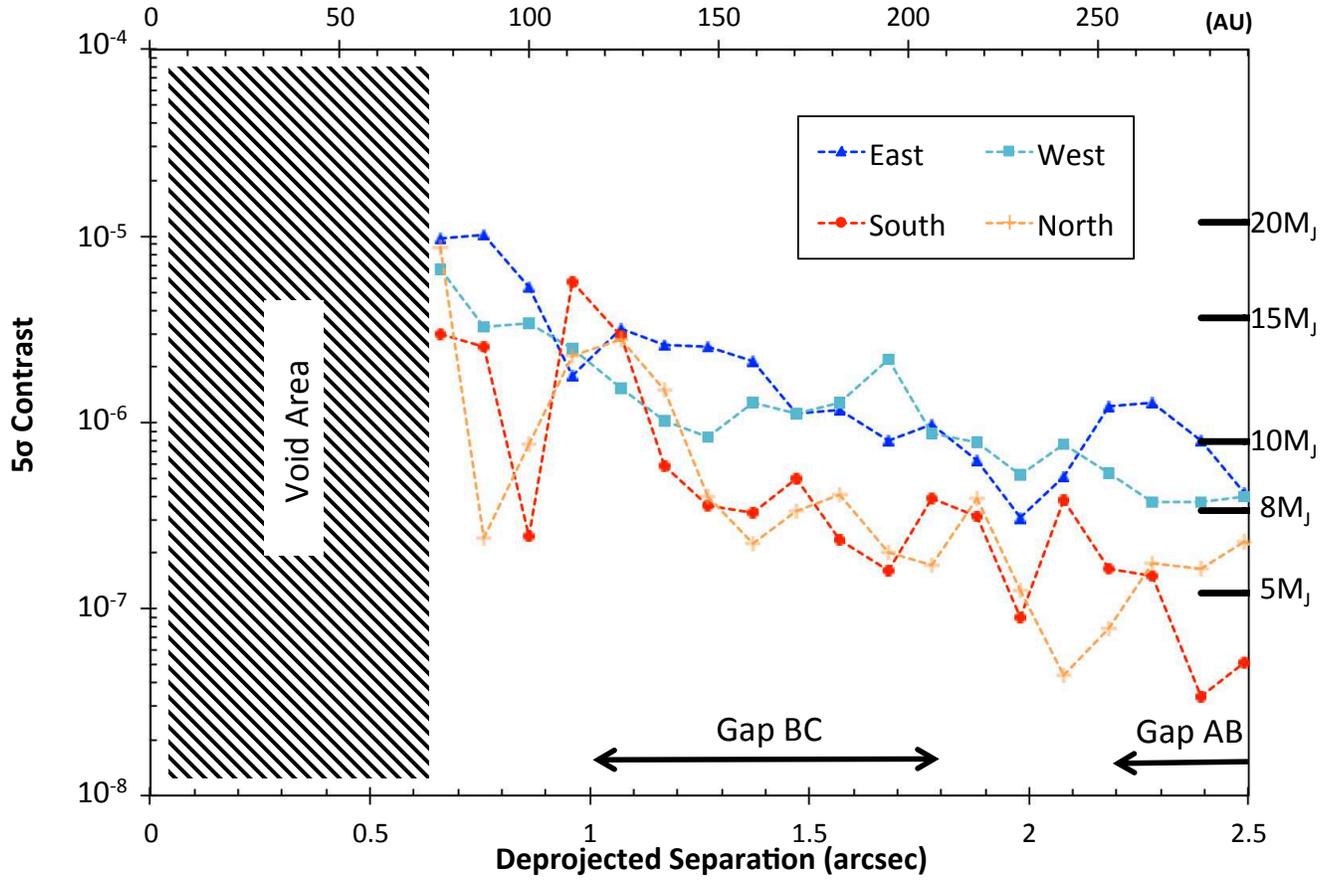}
 \caption{5$\sigma$ detection limits on exoplanets. The abscissa and ordinate axes are deprojected separation and 5$\sigma$ contrast relative to the primary star, respectively. The shaded area is occulted by the wedge. The right ordinate axis shows corresponding mass estimated from the star evolution model (BT-Settl). More detail is described in Section~\ref{sec-dl}.}
 \label{fig-dl}
\end{figure}

We detected no point sources within 2$\arcsec$~($\sim$232~AU) of the star in the PSF-template subtracted image. We calculated the sensitivity to exoplanets that may exist within the disk and report our detection limits (see \citealp{Brandt+McElwain+Turner+etal_2013}). 
The 1$\sigma$ noise was calculated from the standard deviation of SB profile (employed same calculation as Section~\ref{sec-sbsd}) assuming that only the core of the PSF (1~pixel) can be detected. Figure~\ref{fig-dl} shows the 5$\sigma$ (5$\times$1$\sigma$) contrast curve, which we consider the sensitivity limit of the observations. To calculate a contrast relative to the primary, the primary star count rate for our observation was estimated using the STIS imaging ETC\footnote[17]{http://etc.stsci.edu/etc/input/stis/imaging/} since it was occulted by wedge structures. We used a template A0-type spectrum and normalized it using the $V$ magnitude of HD~141569~A. Then the 5$\sigma$ noise was divided by the estimated primary star's count rate (2.1$\times$10$^6$~count~s$^{-1}$). We confirmed the contrast is reasonable by embedding some artificial stars.

Exoplanetary mass estimates are shown in the right side of Figure~\ref{fig-dl}. They were estimated from the BT Settl evolutionary model \citep{Allard+Homeier+Freytag+etal_2011}. We took the spectra of sub-stellar objects (20, 15, 10, 8, and 5~M$_J$ at 5~Myr) from the Phoenix web simulator\footnote[18]{https://phoenix.ens-lyon.fr/simulator/index.faces} and calculated each count rate which would be obtained in our observation, using the STIS imaging ETC. 

In the Gap~BC, which is considered to be formed by a giant planet, the PSF core of the possible planet is from 0.5 to 4.5 counts~s$^{-1}$. It corresponds to the contrast of point sources from 2.1$\times$10$^{-6}$ down to 2.4$\times$10$^{-7}$ in Gap~BC. Converted to mass, 
our luminosity upper limit corresponds to a mass upper limit of 9$\pm$3~M$_J$.

\section{Discussion}
\subsection{Inner disk component}
Compared with previous STIS and ACS imagery, \citet{Mouillet+Lagrange+Augereaus+etal_2001, Clampin+Krist+Ardila+etal_2003} did not recover Component~C due to the occultation. The IWA of ground-base observations have reached 0\farcs3 \citep{Biller+Liu+Rice+etal_2015}, but the sensitivity to faint structures at this IWA are more difficult to recover with their processing because it caused significant self-subtraction at small IWA.

Our imaged Component~C overlaps partially with the disk seen in mid-infrared thermal emission \citep{Fisher+Telesco+Pina+etal_2000,Marsh+Silverstone+Beckelin+etal_2002}. A warm disk in CO emission was detected by \citet{Goto+Usuda+Dullemond+etal_2006} at 13--59~AU based on a more recent stellar parallax from \citet{vanLeeuwen_2007}. 
The Component~C material we now see in scattered light is, in part, in the same stellocentric region as the CO emission. A spatial overlap of especially small-grain dust and gas is expected in a transitional disk. HD~141569~A disk has some transitional disk characteristics, which might indicate that this system is in an evolutionary stage between transitional and debris disks. Our Component~C traces dust beyond the radius reported by \citet{Currie+etal+prep}, and partially overlapping with the warm CO disk reported by \citet{Goto+Usuda+Dullemond+etal_2006} and the PAH disk reported by \citet{Thi+Pinte+Pantin+etal_2015}. 
Small dust grains such as we view in the broadband optical are expected to be coupled to the gas \citep{Takeuchi+Artymowicz_2001}. This prediction can be confirmed with ALMA data (see \citet{Currie+etal+prep}).

\subsection{Dynamical limits on planets}
\citet{Wisdom_1980} used the resonance overlap criterion to analytically derive the relationship between the width of a gap and the mass of the planet in a collisionless disk. \citet{Nesvold+Kuchner_2015} used numerical simulations of a collisional disk to estimate a time-dependent gap law. This collisional gap law is based on gas-free N-body simulations of a planetesimal belt and does not consider the effects of gas drag or radiation pressure on dust dynamics, all of which may be important in the case of HD~141569. However, \citet{Ardila+Lubow+Golimowski+etal_2005} simulated the effects of both gas drag and radiative forces on the dust in HD~141569~A. Therein they found that their simulated planetesimal distribution resembled the observed dust distribution. The gap law can be used to place upper limits on the mass of a planet orbiting in a gap, but it is limited by a degeneracy between the planet mass and the radial distance between the planet and the gap. Because we measure the location of both the inner and outer edge of the gap in the HD~141569~A disk, we can break this degeneracy by assuming that the gap contains a planet on a circular orbit at the midpoint between the gap edges (137.5~AU for Gap~BC). Assuming a system age of 5~Myr and stellar mass of 2.5~M$_{\odot}$ \citep{Weinberger+Rivh+Becklin+etal_2000}, and estimating the vertical optical depth of the disk as $L_{IR}/L_{star} = 1.8\times10^{-2}$ \citep{Merin+Montesinos+Eiroa+etal_2004}, we use the technique of \citet{Nesvold+Kuchner_2015} to place an upper limit of 1~M$_J$ on a hypothetical planet orbiting in Gap~BC.

Because we detect a hot dust population close to the star, we are able to constrain the size of Gap~BC and therefore our upper limit for the planet in Gap~BC, 1~M$_J$, is smaller than those of \citet{Ardila+Lubow+Golimowski+etal_2005} and \citet{Reche+Beust+Augereau_2009}. Further numerical simulations are needed to determine the best-fit system architecture to explain the dust distribution in the disk.

\section{Summary}
We detected an inner disk component around HD~141569~A with HST/STIS in optical. The disk component is traced from 0\farcs4 ($\sim$46~AU) to 1\farcs0 ($\sim$116~AU) after deprojection using i=55\arcdeg. It has SB radial profile breaks, a pericenter offset of 6~AU, a spiral arm at 1\farcs1, and overlaps partially with the previously known warm CO disk.

\acknowledgements
This study is based on observations made with the NASA/ESA HST, obtained at STScI, which is operated by the Association of Universities for Research in Astronomy (AURA), Inc., under NASA contract NAS~5-26555. These observations are associated with program No.13786. Support for program No.13786 was provided by NASA through a grant from STScI. We thank the anonymous referee for helpful suggestions. M.K. acknowledges the support of the NASA exchange program operated by Universities Space Research Association and of the Osaka University Scholarship for Overseas Research Activities 2015. J.C. acknowledges support from the South Carolina Space Grant Consortium REAP Program.

{\it Facilities:} \facility{HST (STIS)}.


\clearpage
\begin{table*}[]
\centering
\caption{Observation log}
\label{table-log}
\begin{tabular}{ccccccccc}
 \hline
 Target Name & UT Date & visit\# & ORIENTAT$^a$ & CORONAGRAPH & EXPTIME$^b$ & NUMEXP$^c$ & TINTTIME$^d$ & DATASET$^e$ \\
 \hline
 HD~141569~A & 2015 Jun 12 & visit1 & 111.53 & BAR5 & 6.3 & 8 & 50.4 & ocjc35010, ocjc35020, ocjc35030 \\
  & & & & WEDGEA1.0 & 400 & 4 & 1600 & ocjc35040\\
  & & visit2 & 95.53 & BAR5 & 6.3 & 8 & 50.4 & ocjc36010, ocjc36020, ocjc36030 \\
  & & & & WEDGEA1.0 & 400 & 4 & 1600 & ocjc36040\\
  & & visit4 & 81.53 & BAR5 & 6.3 & 8 & 50.4 & ocjc38010, ocjc38020, ocjc38030\\
  & & & & WEDGEA1.0 & 400 & 4 & 1600 & ocjc38040\\
 HD~135298 & 2015 Jun 12 & visit3 & 81.01 & BAR5 & 6.8 & 8 & 54.4 & ocjc37010, ocjc37020, ocjc37030\\
  & & & & WEDGEA1.0 & 400 & 4 & 1600 & ocjc37040 \\
 \hline
 HD~141569~A & 2015 August 18 & visit5 & 56.53 & BAR5 & 6.3 & 8 & 50.4 & ocjc31010, ocjc31020, ocjc31030 \\
  & & & & WEDGEA1.0 & 405.7 & 4 & 1622.8 & ocjc31040\\
  & & visit6 & 46.53 & BAR5 & 6.3 & 8 & 50.4 & ocjc32010, ocjc32020, ocjc32030 \\
  & & & & WEDGEA1.0 & 405.7 & 4 & 1622.8 & ocjc32040\\
  & & visit8 & 36.53 & BAR5 & 6.3 & 8 & 50.4 & ocjc34010, ocjc34020, ocjc34030\\
  & & & & WEDGEA1.0 & 405.7 & 4 & 1622.8 & ocjc34040\\
 HD~135298 & 2015 August 18 & visit7 & 58.80 & BAR5 & 6.8 & 8 & 54.4 & ocjc33010, ocjc33020, ocjc33030\\
  & & & & WEDGEA1.0 & 405.5 & 4 & 1622 & ocjc33040 \\
 \hline
 \multicolumn{8}{l}{Note: (a) Position angle of the image $+$Y axis measured eastward from celestial north (degree).}\\
 \multicolumn{8}{l}{(b) Individual exposure times (sec).}\\
 \multicolumn{8}{l}{(c) Number of exposures.}\\
 \multicolumn{8}{l}{(d) Total integration time (sec); $EXPTIME \times NUMEXP$.}\\
 \multicolumn{8}{l}{(e) same as ROOTNAME in the header and fits file name}\\
\end{tabular}
\end{table*}

\clearpage

\clearpage

\clearpage

\clearpage

\end{document}